\documentclass[review,times,fleqn]{elsarticle}
\usepackage[ruled,longend,lined]{algorithm2e}
\usepackage{amsmath}
\usepackage{amssymb}
\usepackage{amsthm}
\usepackage{atbegshi}
\usepackage{graphicx}
\usepackage{hyperref}
\usepackage{makeidx}
\usepackage{mathabx}
\usepackage{multicol}
\usepackage{multirow}
\usepackage{parskip}
\usepackage{rotating}
\usepackage{setspace}
\usepackage{subfig}
\usepackage{times}
\usepackage{ulem}
\usepackage[dvipsnames,svgnames,table]{xcolor}

\newtheorem{theorem}{Theorem}[section]

\newtheorem{lemma}[theorem]{Lemma}

\journal{Pattern Recognition Letters}

\begin{document}

\begin{frontmatter}

\title{Assessing the best edit in perturbation-based iterative refinement algorithms to compute the median string.\footnote{This is an Author’s Original Manuscript of an article whose final and definitive form,
the Version of Record, has been published in Pattern Recognition Letters [copyright Elsevier],
available online at: https://doi.org/10.1016/j.patrec.2019.02.004}}

\author[udec]{P. Mirabal}
\ead{psanchezm@udec.cl}

\author[ucsc]{J. Abreu \corref{cor1}}
\ead{joseabreu@ucsc.cl} \cortext[cor1]{Corresponding author. Fax:965909326 }

\author[udec,mi]{D. Seco}
\ead{dseco@udec.cl}

\address[udec]{Department of Computer Science, Universidad de Concepci\'{o}n, Concepci\'{o}n, Chile}

\address[ucsc]{Departamento de Ingenier\'{i}a Inform\'{a}tica, Universidad Cat\'{o}lica de la Sant\'{i}sima Concepci\'{o}n, Chile}

\address[mi]{Millennium Institute for Foundational Research on Data, Chile}

\begin{abstract}

Strings are a natural representation of biological data such as DNA, RNA and protein sequences. The problem of finding a string that summarizes a set of sequences has direct application in relative compression algorithms for genome and proteome analysis, where reference sequences need to be chosen. Median strings have been used as representatives of a set of strings in different domains. However, several formulations of those problems are NP-Complete. Alternatively, heuristic approaches that iteratively refine an initial coarse solution by applying edit operations have been proposed. Recently, we investigated the selection of the optimal edit operations to speed
up convergence without spoiling the quality of the approximated median string. We propose a novel algorithm that outperforms state of the art heuristic approximations to the median string in terms of convergence speed by estimating the effect of a perturbation in the minimization of the expressions that define the median strings. We present corpus of comparative experiments to validate these results.

\end{abstract}

\begin{keyword}
approximate median string \sep edit distance \sep edit operations
\end{keyword}

\end{frontmatter}

\section{Introduction}\label{sec:intro}

The concept of \textit{median} is useful in many contexts as a representative for a collection of objects.  As defined in \cite{Kohonen1985} the \textit{median} of a set $S$ of strings is the one that minimizes the sum of the distances to each element in the collection. Note that such a string does not need to be part of the set, nor unique. In the case of strings, Levenshtein distance \cite{Levenshtein1966} has been widely used.

For edit distance metrics, \cite{Casacuberta1997} and \cite{Nicolas2005} show that computing the median of a set of strings is a problem within the NP-Complete class for several formulations. Different approximations have therefore been proposed. One of those approaches, called perturbation-based iterative refinement by \cite{Jiang2004} has been studied in \cite{Kohonen1985,Hinarejos2003,Cardenas2004}. The kernel idea is to perform successive edit operations to an initial string while at least one of the perturbations leads to an improvement. Those approaches have proved to converge to quality approximations of the true median \cite{Hinarejos2003}, yet they may require to perform an important number of perturbations before converging. For these methods, it is important to study how to score the goodness of each candidate perturbation in order to test the most promissory ones.

\section{Preliminaries}
Let $\Sigma$ be an alphabet where $\epsilon$ denotes the empty symbol, also $S_{i}$, $S_{i}$ strings over $\Sigma$. An edit operation is a pair $(a,b) \neq (\epsilon, \epsilon)$, written $a \rightarrow b$, which transforms a string $S_{i}$ into $S_{j}$, if $S_{i}=\sigma$a$\tau$ and $S_{j}=\sigma$b$\tau$. Substitutions, deletions and insertions are denoted by $a \rightarrow b$, $a \rightarrow \epsilon$, $\epsilon \rightarrow b$, respectively. Let $E_{S_{i}}^{S_{j}}=\{e_{1}, e_{2}, ..., e_{n}\}$ be a sequence of edit operations transforming $S_{i}$ into $S_{j}$ and, $\omega(a \rightarrow b)$ a function that assigns a cost to an edit operation. The cost of $E$ is $\omega(E)= \sum_{e_i \in E} \omega(e_i)$ and the edit distance from $S_{i}$ to $S_{j}$, is defined as $d(S_{i}, S_{j})=argmin_{E_{S_{i}}^{S_{j}}}\{ \omega(E)\}$.

\section{Related work}\label{sec:RelatedWork}
Results in a seminal work, \cite{Kruskal1983}, allow the computation of the median of a set $S$ of $N$ strings in $\mathcal{O}(l^N)$ for the Levenshtein metric and strings of length $l$. However, the computational burden required makes this approach impractical in most scenarios. Therefore, different heuristics have been proposed to overcome this difficulty with the aim of reducing the search space. 

A general strategy is to build the approximate median, one symbol at a time, from an empty string. A goodness function must be defined in order to decide which symbol is the next to be appended. A greedy implementation is described in \cite{Casacuberta1997}. In \cite{Kruzslicz1999} a tie breaking criterion is presented when many symbols have the same goodness index.

An alternative approach is to build the approximate median by successive refinements of an initial string such as the set median, or the greedy approximation in \cite{Casacuberta1997}.

Starting from the set median, \cite{Kohonen1985} systematically changes the guess string performing insertions, deletions and substitutions in every position. An edition is accepted if it leads to an improvement. One specific order to apply operations is proposed in \cite{Hinarejos2003} and \cite{Hinarejos2003a}. Let $\hat{S}^{t-1}$ be the approximated median at step $t-1$, two possible alternatives are considered in the application of edit operations. The first one, performs substitutions in each position of $\hat{S}^{t-1}$, testing each possible symbol. The new partial solution $\hat{S}^{t}$ is the string, selected from all the new candidates that minimizes $\sum_{S_{i} \in S} d(\hat{S}^{t},S_{i})$. A similar procedure is repeated for deletions, substitutions, and insertions, in such order and until there are no more improvements.  

Another variation is to generate a number of candidate approximations from $\hat{S}^{t-1}$ by separately applying substitutions, deletions and insertions at position $i$. The candidate will be the string with lowest $\sum_{S_{i} \in S} d(\hat{S}^{t},S_{i})$, if it is also lower than $\sum_{S_{i} \in S} d(\hat{S}^{t-1},S_{i})$. If no such as sequence exist, $\hat{S}^{t-1}$ remains as the best approximation found. In the next step, the algorithm repeats the search but from position $i+1$. The search stops when there are not improvements. Theoretical and empirical results demonstrate that this approach achieves good approximations to the true median string.

It is important to note that, to evaluate the goodness of an operation, the approaches discussed above require to compute the distance from the candidate median to strings in the set. No a priori ranking of operations is provided. A core problem is to determine the best operation but with a computational cost lower than the required to compute all the distances from the new candidate to each string. 

An alternative to speed up the computation of the approximated median string is described in \cite{Hinarejos2002}. Some operations are preferred, for example, not all possible substitutions are evaluated. Looking at the cost of substitutions, the two closest symbols to the examined position are selected, and only those symbols tested as viable substitutions. 

Instead of applying operations one by one, other author \cite{Cardenas2004} propose to apply multiple perturbations at once. Results in \cite{Abreu2014} suggest that this approach has a fastest convergence but the quality of the approximated median is worst.

In \cite{Abreu2014} authors take advantage of results in \cite{Bunke2002} to study how to score the goodness of each candidate perturbation in order to test first the most promissory ones. Let $\hat{S}^{t-1}$ be the approximated median at step $t-1$, $e_{k}$ an edit operation and $\hat{S}^{t}$ the string derived from $\hat{S}^{t-1}$ by applying $e_{k}$. Possible edit operations are ranked by $\omega(e_{k})$ as well by the number of strings for whom $e_{k} \in E_{Sk}^{\hat{S^{t-1}}}$.  Results show that the proposed approach drastically improves the convergence speed while preserving the quality of the approximated median in comparison with \cite{Hinarejos2002}. However, whether $\sum_{S_{i} \in S} d(\hat{S}^{t},S_{i}) \leq \sum_{S_{i} \in S} d(\hat{S}^{t-1},S_{i})$ or not is also determined by strings for whom $e_{k} \notin E_{Sk}^{\hat{S^{t-1}}}$, which are not explicitly considered by the authors.  

Our main contribution, described in next section, is an improved heuristic to rank edit operations that shows similar or better results in terms of $\sum_{S_{i} \in S} d(\hat{S},S_{i})$, but demanding less computational effort. 

\section{A new algorithm for computing a quality approximate median string} \label{sec:JRStatistical}

Let $e_{k} \in E_{Si}^{S^{j}}$ be the edit sequence from $S_{i}$ to $S_{j}$ with minimum cost, i.e. $d(S_{i},S_{j})=\omega(E)$, and $S_{i}^{\prime}$ the string derived from $S_{i}$ by the application of $e_{k}$. Results in \cite{Bunke2002} allow to compute $d(S_{i}^{\prime},S_{j})$ as $d(S_{i},S_{j})-\omega(e_{k})$. Thus, since $\omega(e_{k}) \geq 0$, then $d(S_{i}^{\prime},S_{j}) \leq d(S_{i},S_{j})$.

For example, let $\hat{S}^{t-1}$ be the candidate median at step $t-1$ from strings in the set $S=\{S_{1}, S_{2}, S_{3}\}$ and $E^{S^{t-1}}_{S_1}$, $E^{S^{t-1}}_{S_2}$, $E^{S^{t-1}}_{S_3}$ the respective minimum cost edit sequences from $\hat{S}^{t-1}$ to each string in $S$. Suppose there is an edit operation such that $e_{k} \in E^{S^{t-1}}_{S_1} \cap E^{S^{t-1}}_{S_2}$, $e_{k} \notin E^{S^{t-1}}_{S_3}$ and $\hat{S}^{t}$ is derived from $\hat{S}^{t-1}$ by $e_{k}$. As regards of $\sum_{S_{i} \in S} d(\hat{S}^{t},S_{i}) = d(\hat{S}^{t},S_{1})+d(\hat{S}^{t},S_{2})+d(\hat{S}^{t},S_{3})$, applying $e_{k}$ could lead to a minimization since $d(\hat{S}^{t},S_{1}) \leq d(\hat{S}^{t-1},S_{1})$, the same holds for $S_{2}$. However, is also determined by $d(\hat{S}^{t},S_{3})$. 

In the example above, $e_{k}$ splits $S$ into two subsets,  $S^{YES}=\{S_{i} \in S~|~d(\hat{S}^{t},S_{i}) \leq d(\hat{S}^{t-1},S_{i})\}$ and $S^{NO}=S-S^{YES}$, thus $\sum_{S_{i} \in S} d(\hat{S}^{t},S_{i}) = \sum_{S_{i} \in S^{YES}} d(\hat{S}^{t},S_{i}) + \sum_{S_{i} \in S^{NO}} d(\hat{S}^{t},S_{i})$, also $\sum_{S_{i} \in S^{YES}} d(\hat{S}^{t},S_{i}) \leq \sum_{S_{i} \in S^{YES}} d(\hat{S}^{t-1},S_{i})$. 

The heuristic proposed in \cite{Abreu2014} uses this information in order to rank the possible edit operations applicable to $\hat{S}^{t-1}$ to derive $\hat{S}^{t}$, selecting the one that could be expected to lead to the lowest $\sum_{S_{i} \in S} d(\hat{S}^{t},S_{i})$. However, it does not consider how $\sum_{S_{i} \in S^{NO}} d(\hat{S}^{t},S_{i})$ behaves. Also, authors only considered in $S^{YES}$ strings $S_{j}$ for whom $e_{k} \in E^{S^{t-1}}_{Sj}$, which guarantees that $d(\hat{S}^{t},S_{j}) \leq d(\hat{S}^{t-1},S_{j})$. In this work we identify some strings $S_{i}$ that authors assigned to $S^{NO}$ for whom $d(\hat{S}^{t},S_{i}) \leq d(\hat{S}^{t-1},S_{i})$ holds.

\subsection{Heuristic to select the best edit operation}

First, we discuss how to determine an upper bound for $d(\hat{S}^{t},S_{i})$, for some $S_{i} \in S^{NO}$ depending on the type of operation $e_{k}$ and $\omega$ without computing the actual distance. We use this result to propose a new heuristic to select the edit operation that could be expected to minimize $\sum_{S_{i} \in S^{YES}} d(\hat{S}^{t-1},S_{i})$.

\begin{lemma}
\label{lemma:01}	
Let $S_{x}$, $S_{y}$ and $M$ be strings, with $S_{x} \in S$ and $S_{y} \in S$. Also, let $E^{M}_{S_{x}}$ and $E^{M}_{S_{y}}$ be the minimum cost edit sequences transforming $M$ into $S_{x}$ and $S_{y}$ respectively, ${e_{iS_{x}}^{M}} = (a \rightarrow b) \in E_{S_{x}}^{M} $ the operation to be applied to position $i$ of $M$, ${e_{iS_{y}}^{M}} = (a \rightarrow c) \in E_{S_{y}}^{M}$ an edit operation of the same type as ${e_{iS_{x}}^{M}}$, and $\hat{M}$ derived from $M$ by ${e_{iS_{x}}^{M}}$. If $\omega(b \rightarrow c) \leq \omega (a \rightarrow c)$, then $d(\hat{M}, S_{y}) \leq d(M, S_{y})$.
\end{lemma}

\begin{proof}

Let $E^{M}_{S_{y}} = \{ e_{1}^{\prime}, e_{2}^{\prime}, ...,{(a \rightarrow c)}, ... , e_{m}^{\prime}\}$. By definition of the edit distance, deriving $\hat{M}$ from $M$ by ${e_{iS_{x}}^{M}}$ means that $M=\sigma$a$\tau$ and $\hat{M}=\sigma$b$\tau$. Thus, the edit sequence $E^{\hat{M}}_{S_{y}} = \{ e_{1}^{\prime}, e_{2}^{\prime}, ...,{(b \rightarrow c)}, ... , e_{m}^{\prime}\}$ will transform $\hat{M}$ into $S_{y}$. In this case, we can write the cost of those edit sequences as equations (\ref{eq:wMSy}) and (\ref{eq:wRSy}) from where we get that, if $\omega(b \rightarrow c) \leq \omega(a \rightarrow c)$ then $\omega(E^{\hat{M}}_{S_{y}}) \leq \omega(E^{M}_{S_{y}})$.

 \begin{equation}
 \omega(E^{M}_{S_{y}})=\sum_{i=0}^{l-1} \omega(e_{i}^{\prime}) + \omega(a \rightarrow c) + \sum_{i=l+1}^{m} \omega(e_{i}^{\prime}) 
 \label{eq:wMSy}
 \end{equation}

 \begin{equation}
 \omega(E^{\hat{M}}_{S_{y}})=\sum_{i=0}^{l-1} \omega(e_{i}^{\prime}) + \omega(b \rightarrow c) + \sum_{i=l+1}^{m} \omega(e_{i}^{\prime}) 
 \label{eq:wRSy}
 \end{equation}

\end{proof}

As explained before, in \cite{Abreu2014} authors propose two approaches, selecting the $e_{k}$ that maximizes $|S^{YES}|$ where $S^{YES}=\{S_{i} \in S| e_{k} \in E^{\hat{S}^{t-1}}_{S_{i}}\}$. Authors also studied ranking operations by $\omega(e_{k})*|S^{YES}|$. Unlike them, our heuristic uses the result in Lemma \ref{lemma:01} to better asses the goodness of an edit operation. Similar to \cite{Abreu2014}, while computing the distance from $\hat{S}^{t-1}$ to each string $S_{i} \in S$ we track the operation that $E^{\hat{S}^{t-1}}_{S_{i}}$ specifies for the position $p$ of $\hat{S}^{t-1}$. Without loss of generality, let us suppose that we have three operations for that position,  $e_{1} \in E^{\hat{S}^{t-1}}_{S_{1}}$, $e^{2} \in E^{\hat{S}^{t-1}}_{S_{2}}$ and  $e_{3} \in E^{\hat{S}^{t-1}}_{S_{3}}$. We know that, if we apply $e_{1}$, $d(\hat{S}^{t-1}, S_{1})$ will decrease. However, we can also consider its \textit{repercussion}\footnote{Hence the name of our method.} over $d(\hat{S}^{t-1}, S_{2})$ and $d(\hat{S}^{t-1}, S_{3})$ if conditions of lemma \ref{lemma:01} hold. We rank operations by  $\omega(e_{k})*|S^{YES}|$ but defining $S^{YES}$ different to \cite{Abreu2014}. In our case $S^{YES}=S^{\prime YES} \cup  S^{\prime \prime YES}$ where $S^{\prime YES}=\{S_{i}|e_{k} \in E^{\hat{S}^{t-1}}_{S_{i}}\}$ and $S^{\prime \prime YES}=\{S_{j}|S_{j} \notin S^{\prime YES}$ and $(\ref{lemma:01})$ holds $\}$

\subsection{An illustrative example}

Let us consider an alphabet $\Sigma = \{ 0, 1, 2, 4,  \epsilon \}$ and a substitution function with the costs in the following matrix.:

\begin{table}[h]
\centering

\label{my-label}
\begin{tabular}{llllll}
\textbf{}                       & \textbf{0}             & \textbf{1}             & \textbf{2}             & \textbf{4}             & $\epsilon$               \\ \cline{2-6} 
\multicolumn{1}{l|}{\textbf{0}} & \multicolumn{1}{l|}{-} & \multicolumn{1}{l|}{1} & \multicolumn{1}{l|}{2} & \multicolumn{1}{l|}{4} & \multicolumn{1}{l|}{2} \\ \cline{2-6} 
\multicolumn{1}{l|}{\textbf{1}} & \multicolumn{1}{l|}{1} & \multicolumn{1}{l|}{-} & \multicolumn{1}{l|}{1} & \multicolumn{1}{l|}{3} & \multicolumn{1}{l|}{2} \\ \cline{2-6} 
\multicolumn{1}{l|}{\textbf{2}} & \multicolumn{1}{l|}{2} & \multicolumn{1}{l|}{1} & \multicolumn{1}{l|}{-} & \multicolumn{1}{l|}{2} & \multicolumn{1}{l|}{2} \\ \cline{2-6} 
\multicolumn{1}{l|}{\textbf{4}} & \multicolumn{1}{l|}{4} & \multicolumn{1}{l|}{3} & \multicolumn{1}{l|}{2} & \multicolumn{1}{l|}{-} & \multicolumn{1}{l|}{2} \\ \cline{2-6} 
\multicolumn{1}{l|}{$\epsilon$}   & \multicolumn{1}{l|}{2} & \multicolumn{1}{l|}{2} & \multicolumn{1}{l|}{2} & \multicolumn{1}{l|}{2} & \multicolumn{1}{l|}{-} \\ \cline{2-6} 
\end{tabular}
\caption{Cost of Substitutions}
\end{table}

Our current candidate string is $\hat{S}^{t-1} = "2"$. The strings in the set are $S_1="0"$, $S_2="1"$ and $S_3="4"$, $\omega(E_{\hat{S}^{t-1}_{S_1}})= \omega(2 \rightarrow 0) = 2$, $\omega(E_{\hat{S}^{t-1}_{S_2}})= \omega(2 \rightarrow 1) = 1$ and  $\omega(E_{\hat{S}^{t-1}_{S_3}})= \omega(2 \rightarrow 4) = 2$. Now $\sum_{S_{i} \in S}
d(\hat{S}^{t-1},S_{i}) = 5$, and we have to  find $\sum_{S_{i} \in S}
d(\hat{S}^{t},S_{i}) < 5$. We can try three possible perturbations for $\hat{S}^{t}$: $(2 \rightarrow 0)$, $(2 \rightarrow 1)$ and $(2 \rightarrow 4)$. If $S$ is divided in $S^{\textsc{YES}}$ and $S^{\textsc{NO}}$ depending only in whether an operation is in the transformation sequence we get this in Table \ref{table:syessno}.

\begin{table}[ht!]
\centering
\begin{tabular}{|c|c|c|c|c|}
\hline
              & \multicolumn{2}{c|}{Pessimistic Analysis}          & \multicolumn{2}{c|}{Repercussion Analysis}          \\ \hline
Op & $S^{\textsc{YES}}$ & $S^{\textsc{No}}$ & $S^{\textsc{YES}}$ & $S^{\textsc{No}}$ \\ \hline
$Op_1$        & $S_1$                         & $S_2,S_3$                    & $S_1$                         & $S_2,S_3$                    \\ \hline
$Op_2$        & $S_2$                         & $S_1, S_3$                   & $S_1, S_2$                    & $S_3$                        \\ \hline
$Op_3$        & $S_3$                         & $S_1, S_2$                   & $S_3$                         & $S_1, S_2$                   \\ \hline
\end{tabular}
\caption{Pessimistic Analysis vs Repercussion Analysis}
\label{table:syessno}
\end{table}

In this case, $(2 \rightarrow 0)$ and $(2 \rightarrow 4)$ are tied, because both will directly decrease distance in 2, but the repercussion in the rest of $S$ is not considered. However, taking a closer look at the substitution matrix, we can see that if $(2 \rightarrow 1)$  is applied, both the distances to $S_1$ and to $S_2$ will decrease. Then, we want to check how one operation affects the rest. We summarize the analysis in Table \ref{table:trep}. We can see in the last column that $(2 \rightarrow 1)$ is the best possible operation, reducing the total distance in 1.

\begin{table*}[ht!]
\centering
\begin{tabular}{|p{1.2cm}|c|c|c|c|p{.5cm}|p{.6cm}|}
\hline
\multirow{2}{*}{Op} & \multirow{2}{*}{\begin{tabular}[c]{@{}c@{}}Direct effect over\\ $\sum_{S_{i} \in S}d(\hat{S}^{t-1},S_{i})$\end{tabular}} & \multicolumn{3}{c|}{\begin{tabular}[c]{@{}c@{}}Indirect effect over \\ $\sum_{S_{i} \in S}d(\hat{S}^{t-1},S_{i})$\end{tabular}} &              &       \\ \cline{3-7} 
                    &                                                                                                                          & $(2 \rightarrow 0)$                       & $(2 \rightarrow 1)$                      & $(2 \rightarrow 4)$                      & Rep. & Total \\ \hline
$(2 \rightarrow 0)$ & 2                                                                                                                        & 0                                         & 0                                        & -2                                       & -2           & 0     \\ \hline
$(2 \rightarrow 1)$ & 1                                                                                                                        & 1                                         & 0                                        & -1                                       & 0            & 1     \\ \hline
$(2 \rightarrow 4)$ & 2                                                                                                                        & -2                                        & -1                                       & 0                                        & -3           & -1    \\ \hline
\end{tabular}
\caption{Repercussion costs and new ranking}
\label{table:trep}
\end{table*}

\subsection{Computing the approximate median string}

Algorithm~\ref{algo:AppMedianStringRepercussion(S,R)} shows our method to compute the approximate median string.

\begin{algorithm}[ht!]
    \SetKwInOut{Input}{Input}
    \SetKwInOut{Output}{Output}
    \SetKwFor{PFor}{parfor}{do}{end}
    \LinesNumbered
    \DontPrintSemicolon
    \SetVlineSkip{0.5ex}
    \SetCommentSty{textit}
    \vspace{0pt}
      \small
      \Input{instance set $S$, initialization string $R$ }
      \Output{approximate median string $\hat{S}$}
      \BlankLine
$R^{\prime}=R$\;
\Repeat{no operation $op_i$ applied to $\hat{S}$ improve the result}
{
$\hat{S}=R^{\prime}$\;
\ForEach{instance $s_i \in S$}
{
compute $D(R^{\prime},s_i)$\;
obtain that $Q^{R^{\prime}}_{si}$ is the minimum cost edit sequence needed to transform $R^{\prime}$ into $s_i$\;
update statistics for the operation in each position $j$ of $R^{\prime}$\;
}
\ForEach{position $j \in R^{\prime}$}
{
	\ForEach{symbol si $\in \Sigma$   } 
	{
			\ForEach{symbol sj $\in \Sigma$   }
			{
				$Rep[j].subst[si] += SubstCost(R[j], si) - SubstCost(si, sj)\;$
				$Rep[j].ins[si] += InsCost(sj) - SubstCost(si, sj)\;$
			}
	}
}
\While{$\sum_{s_i \in S} D(\hat{S},s_i) \leq \sum_{s_i \in S} D(R^{\prime},s_i)$ {\bf and} $O_p \neq \emptyset$}
{
  $op_i=O_{p}.dequeue$\;
	obtain a new candidate $R^{\prime}$ applying $op_i$ to $\hat{S}$\;
}
}
\Return $\hat{S}$\;
      \vspace{1ex}
      \caption{AppMedianStringRepercussion(S,R) :$\hat{S}$}
      \label{algo:AppMedianStringRepercussion(S,R)}
\end{algorithm}
    
This algorithm is based on the {\tt AppMedianString} function presented in~\cite{Abreu2014}, being the main differences in lines 9-15. The original algorithm stores all possible operations in a priority queue $O_p$ sorted by goodness index, which is based on the statistics computed in lines 4-8. In lines 9-15, we collect the possible repercussion of each operation on the elements in $S^{NO}$, which is used to compute a better goodness index.

This algorithm iterates by considering one permutation at a time, until it does not get any improvement during the iteration. Each iteration may consider several different operations. In the worst case, this is upper bounded by $O(l \times \Sigma^2)$, where $l$ is the length of the longest string. In the experimental evaluation, we show that this bound is rather pessimistic and our heuristic usually needs just a few operations per iteration. This is a key difference with the algorithm in~\cite{Abreu2014}, which uses more operations per iteration. For each operation explored during an iteration, the algorithm computes the distance of the new candidate $R^{\prime}$ to all the elements in $S$ (lines 16-19), which takes time $O(N \times dc)$, where $dc$ is the time to compute the edit distance and depends on the specific measure used. By providing a better ranking, we save on the number of operations explored per iteration, and thus, on the number of times this distance is computed, which is expensive. However, to do that, we expend some computations to bound the \textit{repercussion} (lines 9-15). This computation takes $O(l \times \Sigma^2)$ time and it is usually worth it as $l$ and $\Sigma$ are much lower than $N$ in most applications.
\section{Experimental results}\label{sec:exp}

We use two datasets. The first one corresponds to the directions of Freeman chain codes \cite{Freeman1974}, where alphabet is $\Sigma=\{0,1,2,3,4,5,6,7,\epsilon\}$ and $\epsilon$ denotes the empty symbol used for deletions and insertions. The strings in this dataset represent contours of letters from a widely known 2D shape database, the \textit{NIST-3 Uppercase Letters}, \cite{Jain1997, Garcia-Diez2011, Rico2012}, with $26$ classes, one for each letter of the English alphabet. Each class is composed by $360$ samples. In addition, we generated an artificial dataset of proteins, with alphabet size of 23 and composed by $720$ samples. The average length of the samples is $500$. 

We compare three algorithms: The current state of the art, labeled as \textit{Frequency*Cost}, one variation that only considers Frequency labeled \textit{Frequency}, and our proposal, labeled \textit{Repercussion}. The first two are described in~\cite{Abreu2014}.

\begin{figure*}[ht!]
 \centering
  \subfloat[Operations vs Size of the Set]{
   \label{fig:OQLetters}
    \includegraphics[width=0.45\textwidth]{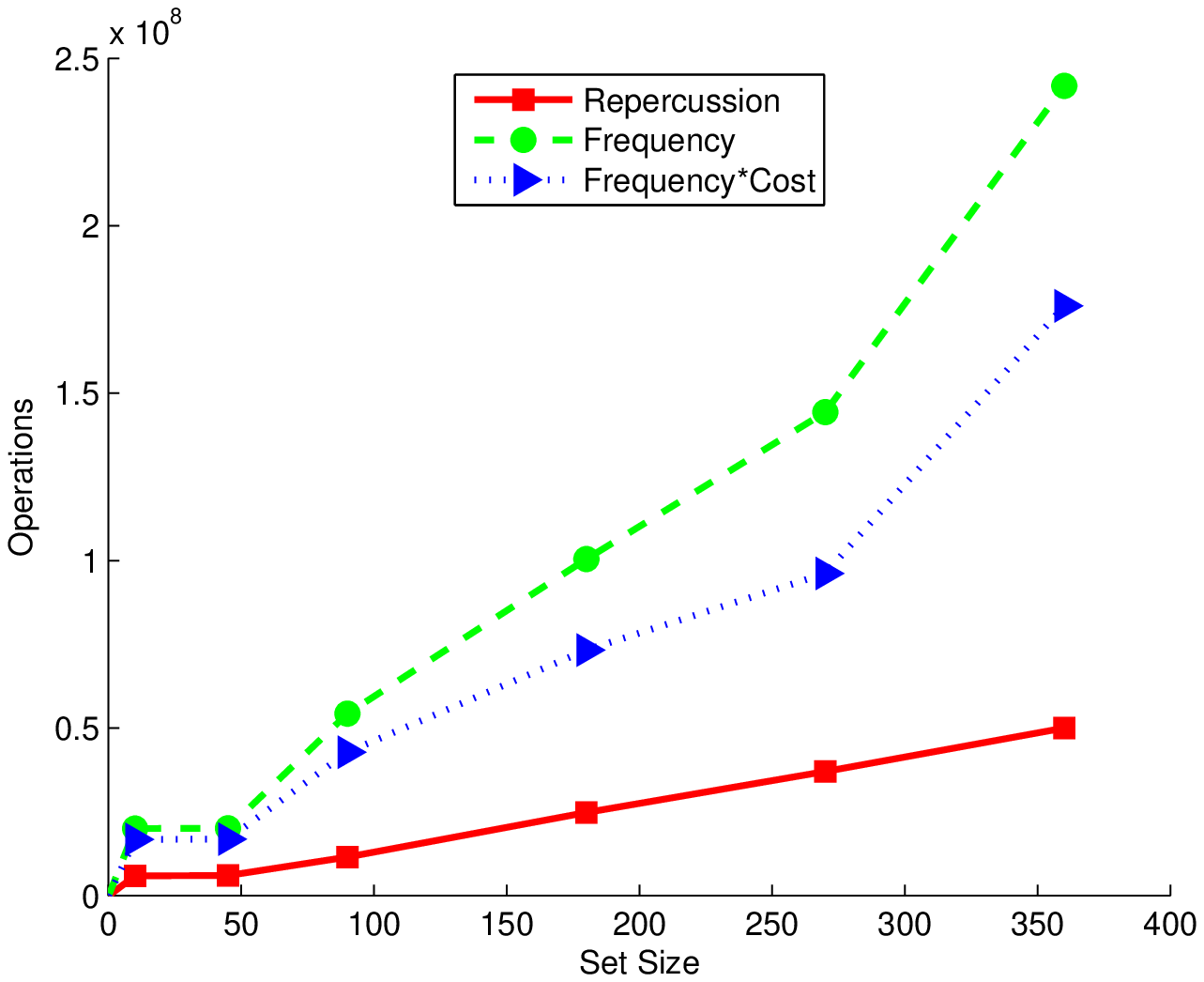}}
  \subfloat[Median Quality]{
   \label{fig:CMLetters}
    \includegraphics[width=0.45\textwidth]{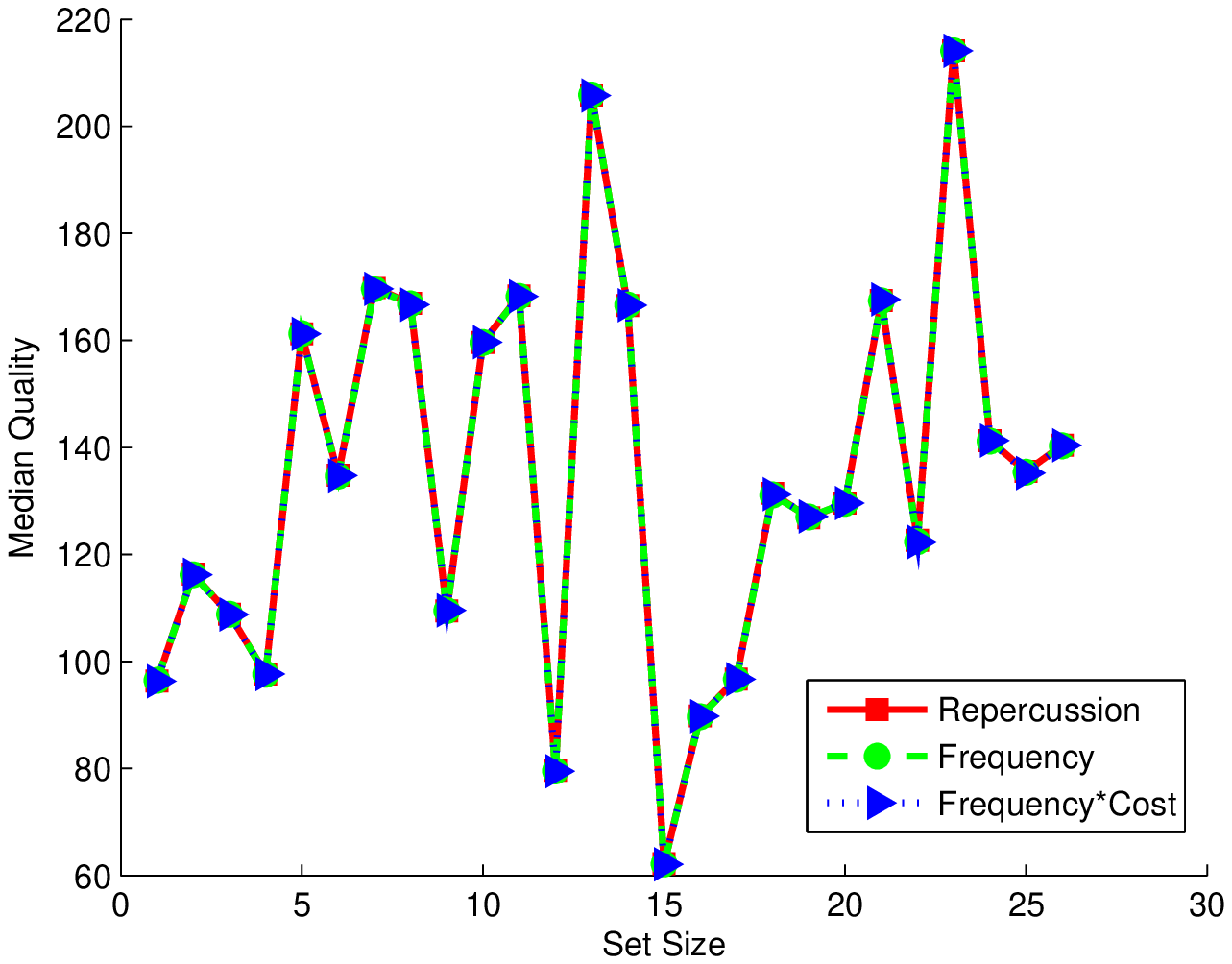}}
  \caption[Letters Dataset Results]{Letters Dataset Results}
 \label{f:animales}
\end{figure*}

\begin{figure*}[ht!]
 \centering
  \subfloat[Operations vs Size of the Set]{
   \label{fig:OQProteins}
    \includegraphics[width=0.45\textwidth]{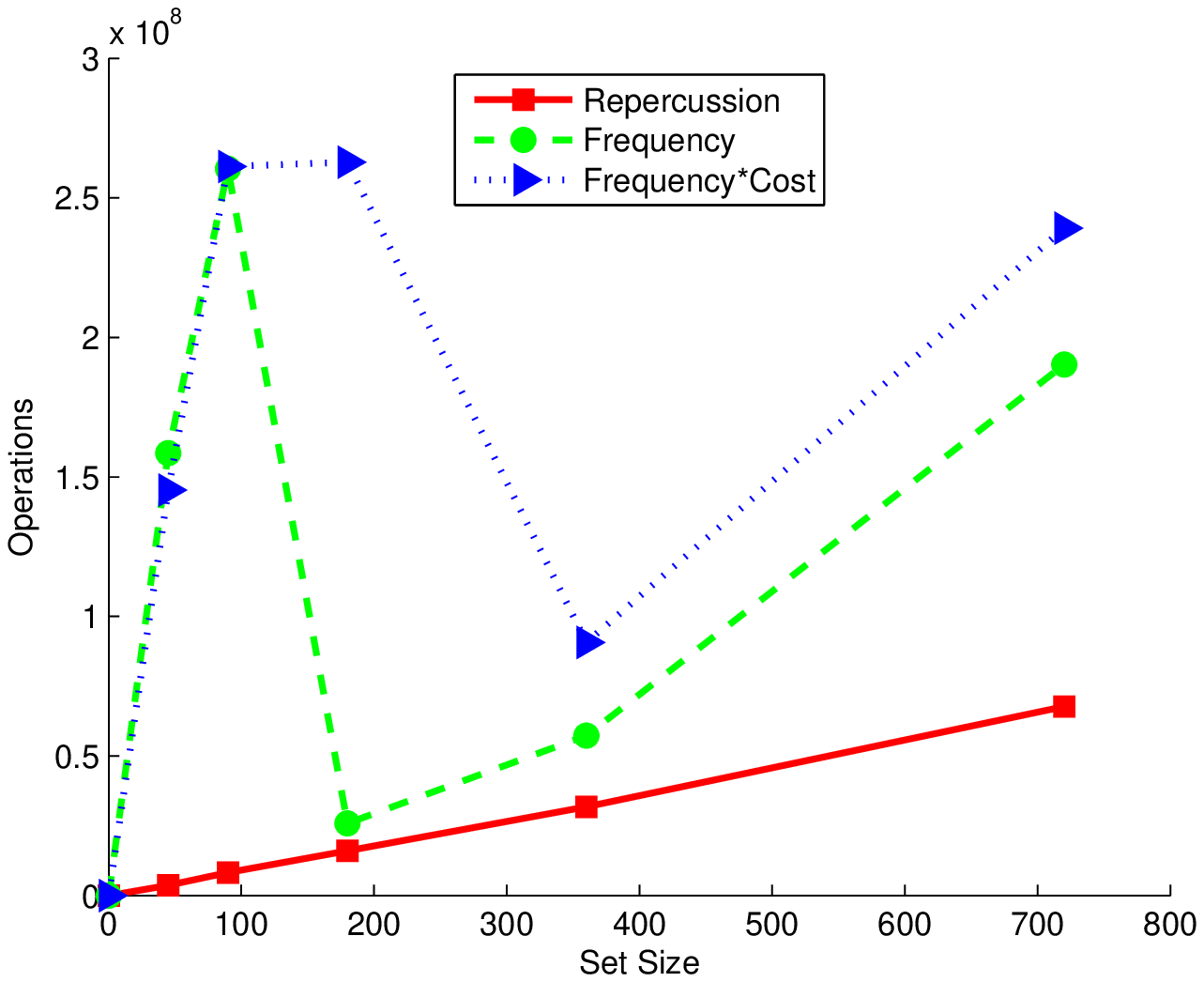}}
  \subfloat[Median Quality]{
   \label{fig:CMProteins}
    \includegraphics[width=0.45\textwidth]{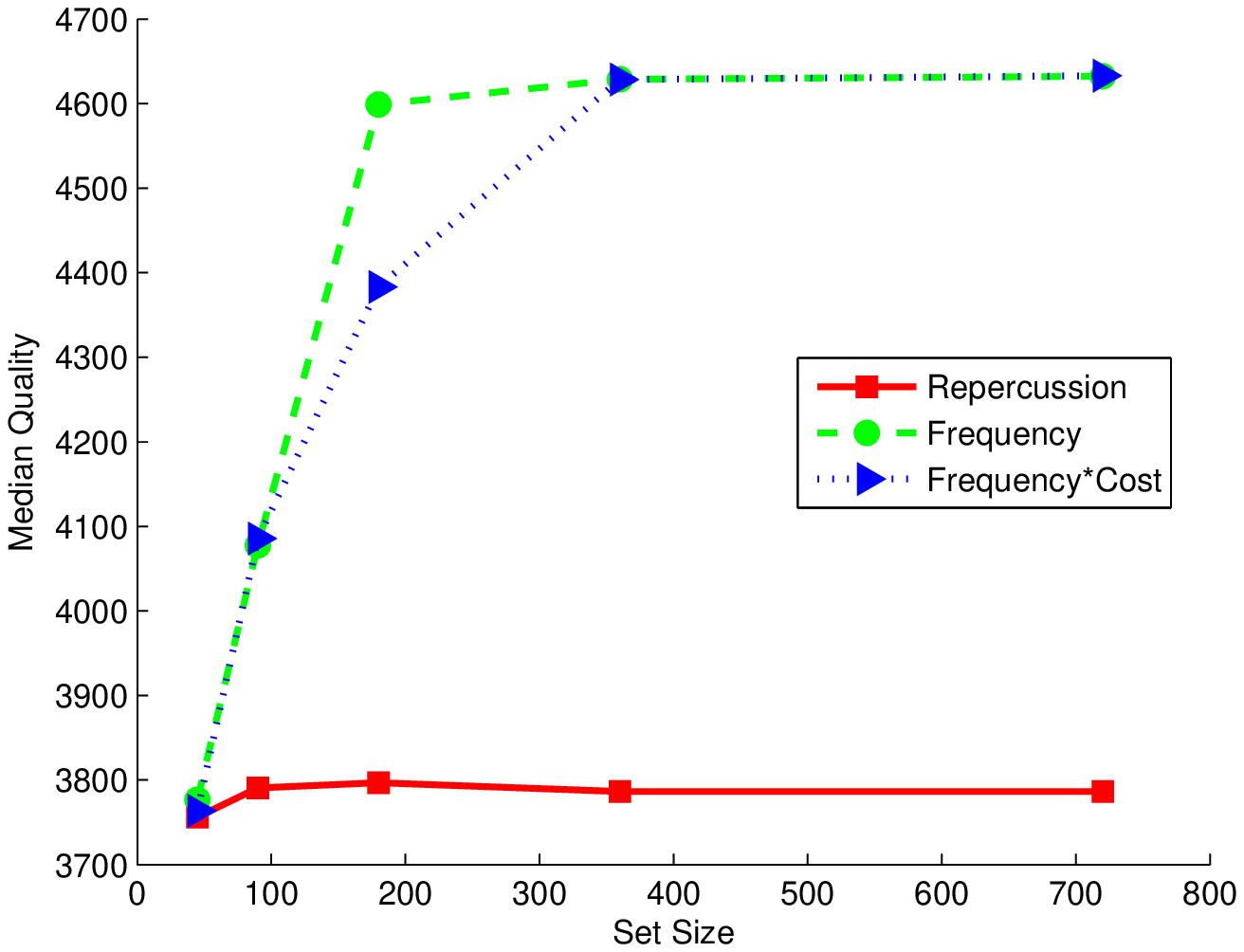}}
  \caption[Proteins Dataset Results]{Proteins Dataset Results}
\end{figure*}

Fig. \ref{fig:OQLetters} and Fig. \ref{fig:OQProteins} show that the number of operations required by the algorithms to converge is significantly lower in our proposal for both datasets. Also important, our heuristic scales better with the size of the dataset. It is worth noting that we measure number of operations, including those used to compute statistics and repercussion.

As a sanity check, in Fig. \ref{fig:CMLetters} we show that the quality of the median remains statistically equivalent in the Letters Dataset. However, in the Proteins Dataset we obtained a surprising result, as our proposal achieves a better median than the state of the art, as we can see in  Fig. \ref{fig:CMProteins}. This means that the other methods get stuck in a local optimum. Future research may compare these and other methods in terms of robustness to local optimums.

\begin{figure}[ht!]
 \centering
    \includegraphics[width=0.45\textwidth]{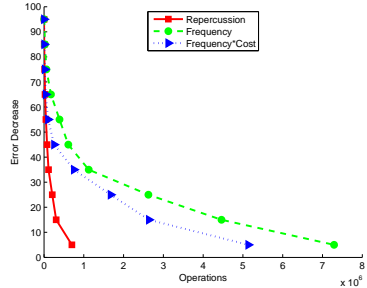}
 \caption[Error Decrease in Letters Dataset]{Error Decrease in Letters Dataset}
 \label{fig:ED46}
\end{figure}

\begin{figure}
 \centering
    \includegraphics[width=0.45\textwidth]{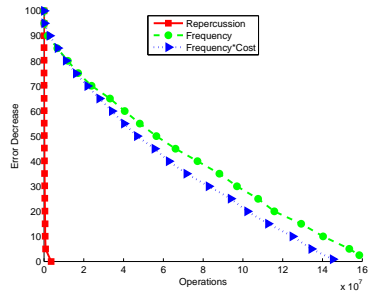}
  \caption[Error Decrease in Proteins Dataset]{Error Decrease in Proteins Dataset}
 \label{fig:ED361}
\end{figure}

Finally, we study how the average distance from the candidate median to each string in the set decreases with the number of iterations. We call this magnitude the error decrease. This prevents misleading conclusions in cases of algorithms which results are obtained by a stagnation (or by a drastically decrease) in the last steps. In other words, it is desirable to provide algorithms that, not only require few operations to converge, but also their convergence speed is fast. Fig. \ref{fig:ED46} and Fig. \ref{fig:ED361} show that our method has a great convergence speed.

\section{Conclusions and Future work}\label{sec:conclusions}

A new approach to compute a quality approximation to the
median string has been presented. The algorithm builds an approximate
median through the successive refinements of a partial solution.
Modifications are applied one by one in a manner similar to
that of Martínez-Hinarejos et al. (2003), and empirical results show
that this approach leads to better approximations than those
methods which apply several perturbations simultaneously,
although the latter runs much faster. Comparisons with
Martínez-Hinarejos (2003) show that the proposed algorithm is
able to compute high-quality approximations to the true median
string but requires significantly less computation and is about 10
times faster, which makes it highly suitable for applications that
require a precise approximation. As pointed out in Section 2, an
operation opi determines two subsets SYES and SNO from S. Applying
opi to bS results in new string bS0 such as the distance from strings in
SYES to bS0 will decrease. Further research may address to better
characterize how the distance from bS0 to strings in S
NO behaves
without computing those distances, but using information gathered
when computing the distances to bS. This can help to select
the best operation to reduce the number of distances computed
without spoiling the approximation quality. Another subject of
interest is to analyse how the choice of a different optimal path
will affect results, since a different ranking might be obtained.

\section*{Acknowledgements}

This paper is funded in part by CONICYT through a Ph.D. Scholarship number $63140074$; the Universidad Católica de la Santísima Concepción through the research project DIN-01/2016; European Union's Horizon 2020 research and innovation programme under the Marie Sk\l odowska-Curie grant agreement 690941; Millennium Institute for Foundational Research on Data (IMFD); 
and Fondecyt-Conicyt grant number 1170497.



\begin{thebibliography}{18}

\bibitem{Abreu2014}J. Abreu and J.R. Rico-Juan. A new iterative algorithm for computing a quality approximate median of strings based on edit operations. Pattern Recognition Letters, 36:74 – 80, 2014.

\bibitem{Bunke2002} Horst Bunke, Xiaoyi Jiang, Karin Abegglen, and Abraham Kandel. On the weighted mean of a pair of strings. Pattern Analysis and Applications, 5:23–30, 2002. 10.1007/s100440200003.

\bibitem{Cardenas2004}R. A. Mollineda Cardenas. A learning model for multiple-prototype classification of strings. In 17th Int. Conf. on Pattern Recognition (ICPR), volume 4, pages 420–423, 2004.

\bibitem{Casacuberta1997}Francisco Casacuberta and M.D. Antonio. A greedy algorithm for computing approximate median strings. VII Simposium Nacional de Reconocimiento de Formas y Analisis de Imagenes, pages 193–198, 1997.

\bibitem{Fisher2000}Igor Fischer and Andreas Zell. String averages and self-organizing map for strings. In Proceedings of the Neural Computation 2000, Canada / Switzerland, ICSC, pages 208–215. Academic Press, 2000.

\bibitem{Freeman1974}H. Freeman. Computer processing of line-drawing data. Computer Surveys, 6:57–96, 1974.

\bibitem{Garcia-Diez2011}Silvia Garc\'{i}a-D\'{i}ez, Franois Fouss, Masashi Shimbo, and Marco Saerens. A sumover-paths extension of edit distances accounting for all sequence alignments. Pattern Recognition, 44(6):1172 – 1182, 2011.

\bibitem{Jain1997}A.K. Jain and D. Zongker. Representation and recognition of handwritten digits using deformable templates. IEEE Trans. on Pattern Analysis and Machine Intelligence, 19:1386 –1390, 1997.

\bibitem{Jiang2004}X Jiang, H Bunke, and J Csirik. Median strings: A review. In M. Last, A. Kandel, and H. Bunke, editors, Data Mining in Time Series Databases, pages 173–192. World Scientific, 2004.

\bibitem{Kohonen1985}T. Kohonen. Median strings. Pattern Recognition Letters, 3:309–313, 1985. 

\bibitem{Kruskal1983}J. Kruskal. An overview of sequence comparison. time warps, string edits and macromolecules. SIAM Reviews, 2:201–2037, 1983.

\bibitem{Kruzslicz1999}F. Kruzslicz. Improved greedy algorithm for computing approximate median strings. Acta Cybernetica, 14:331–339, 1999.

\bibitem{Levenshtein1966}Vladimir I Levenshtein. Binary codes capable of correcting deletions, insertions and reversals. In Soviet physics doklady, volume 10, pages 707–710, 1966.

\bibitem{Hinarejos2002}Carlos D. Mart\'{i}nez-Hinarejos, Alfonso Juan, Francisco Casacuberta, and Ramon Mollineda. Structural, Syntactic, and Statistical Pattern Recognition: Joint IAPR 13International Workshops SSPR 2002 and SPR 2002 Windsor, Ontario, Canada, August 6–9

\bibitem{Hinarejos2003}Carlos David Mart\'{i}nez-Hinarejos. La cadena media y su aplicacion ´ en reconocimiento de formas. PhD thesis, Universidad Politcnica de Valencia, 2003.

\bibitem{Hinarejos2003a}C.D. Mart\'{i}nez-Hinarejos, A. Juan, and F. Casacuberta. Median strings for knearest neighbour classification. Pattern Recognition Letters, 24(1-3):173–181, 2003.

\bibitem{Nicolas2005}Francois Nicolas and Eric Rivals. Hardness results for the center and median string problems under the weighted and unweighted edit distances. Journal of Discrete Algorithms, 3(2-4):390–415, 2005. Combinatorial Pattern Matching (CPM) Special Issue. The 14th annual Symposium on combinatorial Pattern Matching.

\bibitem{Rico2012}Juan Ramon Rico-Juan and Jos ´ e Manuel I nesta. New rank methods for reducing the size of the training set using the nearest neighbor rule. Pattern Recognition Letters, 33(5)(5):654 – 660, 2012.
\end{thebibliography}



\end{document}